\documentclass[twocolumn]{jpsj3}
\usepackage{bm,amsmath,amssymb,stmaryrd} 
\usepackage{graphicx}
\usepackage{txfonts}

\renewcommand{\Im}{\mathop{\mathrm{Im}}\nolimits} 
\newcommand{\affiliation}[1]{\inst{#1}}  

\begin{document}

\title{Spin Current Generation as a
  Nonequilibrium Kondo Effect in a Spin-orbit Mesoscopic Interferometer}
 
\author{Nobuhiko Taniguchi and Kenta Isozaki}

\affiliation{Institute of Physics, University of Tsukuba, Tennodai
  Tsukuba 305-8571, Japan}

\date{\today}

\abst{%
  We study nonequilibrium generation of spin-dependent transport
  through a single-level quantum dot embedded in a ring with the
  Rashba spin-orbit coupling. We consider nonmagnetic systems,
  involving no magnetic field nor ferromagnetic leads.  It is
  theoretically predicted that large spin-dependent current occurs as
  a combined effect of the Rashba spin-orbit interaction, the Kondo
  effect, and nonequilibrium effect, without using magnetic field or
  material.  The phenomenon is viewed as a new nonequilibrium
  correlation effect that disappears when either interaction or finite
  bias is absent. We show how the Kondo physics is connected with such
  emergent spin phenomenon by employing the finite interaction
  slave-boson approach.}


\kword{spin-orbit interaction, quantum dot, Kondo effect, Anderson
  impurity model}

\maketitle%

\section{Introduction}

In semiconducting devices, the Rashba spin-orbit (SO) interaction is
generated by the potential asymmetry in the direction perpendicular to
the semiconductor plane. By utilizing tunable spin-orbit interaction, there
has been growing interest in manipulating electron spins toward
realizing controllable semiconductor spintronics devices nonmagnetically.
In coherent transport through a nanostructure or a quantum dot, the
Coulomb interaction plays a fundamental role. 
Among such striking
many-body effects is the Kondo effect of semiconductor quantum
dots~\cite{Ng88, Goldhaber-Gordon98,Cronenwett98}.  
Whereas the dot blockades the coherent channel by its repulsive
interaction on the dot (the Coulomb blockade effect), decreasing
temperature helps develop a strong singlet correlation between
electrons on the dot and in the leads, showing conductance enhancement
(the Kondo effect).  The phenomenon is characterized by the Kondo
temperature, which is also controllable in semiconductor
nanostructures.
Such a strongly coupled quantum dot is known to produce large spin
polarized currents once magnetic field applies~\cite{Costi01}.  We
explore the possibility of spin transport of a similar nature
\emph{nonmagnetically}, being maintained electrically by the
spin-orbit interaction.

Controllable quantum interference effect enables us to regulate
coherent charge and spin transport.
Since one may describe the effect of Rashba spin-orbit interaction
effectively as a spin-dependent phase $\phi_{\sigma} = \sigma
\phi_{\text{so}}$ (with $\sigma = \pm 1$) for an
interferometer~\cite{Sun05a}, the Rashba spin-orbit interferometer (SOI)
shares much similarity with the Aharonov-Bohm interferometer
(ABI) where the Aharonov-Bohm (AB) phase $\phi_{\text{AB}}$ is
spin-independent.
In the ABI, the conductance enhancement due to the Kondo effect is
easily deformed by quantum interference by a direct hopping channel
(the Fano-Kondo effect~\cite{Hofstetter01,Bulka01,Kim03,Aharony05,Takahashi06});
Charge current can be controlled by adjusting gate voltage, the AB
flux, and bias voltage.  We expect similar controllability is
attainable on spin current in the SOI.

Several kinds of suggestions have been made to realize spin transport
by using the Rashba spin-orbit coupling.  The two systems (SOI and ABI)
produce exactly the same \emph{linear} conductance (and no
spin-dependent conductance), because $\phi_{\sigma}$ appears only in
the form of $\cos\phi_{\sigma}$ in linear transport.  Hence one should
go into a nonequilibrium regime with finite bias voltage to
realize spin-dependent phenomena.  Alternatively one uses magnetic
systems such as systems with ferromagnetic leads or systems with both
the AB and the Rashba fluxes.
Indeed, the formation of spin moment was studied in nonequilibrium
Rashba dots~\cite{Sun06,Crisan09}; Spin-charge filtering was proposed
by combining with the AB effect~\cite{Heary08}; bias-induced
generation of spin polarization current was predicted~\cite{Lu07}.
Particularly challenging is the possibility to generate and regulate
spin transport only via electric field in nonmagnetic systems.  Yet,
electrically emergent spin transport is still a largely unexplored
phenomena in a nonequilibrium many-body system.  It is crucial to
understand whether and how the SO interaction and strong correlation
effect such as the Kondo physics are responsible for it.


In this paper, we investigate the nonequilibrium electric generation
of spin transport in a mesoscopic spin-orbit interferometer.  The
system we consider is nonmagnetic, involving no magnetic field nor
ferromagnetic leads.  We are particularly interested in how spin
transport can be spontaneously realized in a ring geometry as a result
of combined effect of nonequilibrium nature (finite bias voltage), the
Rashba SO interaction, and many-body effect (the Kondo physics).
We employ the finite interaction slave-boson approach on the
single-level Anderson model with the SO interaction and direct hopping
between leads.  The approximation is valid up to the order of the 
Kondo temperature, beyond which the Kondo effect is suppressed.
Within its validity, we will find that spin transport through a
single-level quantum dot occurs in the Kondo valley if all the
following three conditions are met: (1) strong Coulomb interaction is
present on the dot, (2) under finite bias voltage where $eV$ is
roughly on the order of the Kondo temperature or less, (3) at
temperature lower than the Kondo temperature.
It will also be confirmed that either too large bias voltage or
temperature (comparing to the Kondo characteristic temperature)
destroys this spin-dependent transport along with suppressing the
Kondo effect.


In recognizing emergent phenomena as related to the Kondo physics, we
need to know a concrete knowledge of the energy scale characterizing
the Kondo physics of the system in the presence of $\phi_{\sigma}$,
which we designate as $T^{*}$ for the sake of clarity; it is
nothing but the standard Kondo temperature for the single
impurity Anderson model (SIAM).
The Kondo temperature represents a crossover (not a
transition) separating the weak-coupling and strong-coupling regions,
the definition has some ambiguity on principle.
Moreover, as for the ABI system, the presence of the AB flux is known
to affect the Kondo temperature
considerably~\cite{Simon05,Lewenkopf05,Yoshii08}; the same goes for a
system with the SO interaction.
To resolve the issue of ambiguity, we resort to the idea of universal
scaling of conductance on the temperature \cite{Costi94} or on the
bias voltage~\cite{Rincon09} in the Kondo regime.


The paper is organized as follows.  In Sec.~\ref{sec:theory}, we
introduce the model of quantum transport of a mesoscopic spin-orbit
interferometer.  The exact formulae of charge and spin currents are
presented in terms of nonequilibrium Green functions, and our choice
of the approach, the finite interaction slave-boson mean field theory,
is briefly summarized.  Section~\ref{sec:spin-moment} clarifies how
the spin-dependent phase induces finite spin density on a
noninteracting or interacting dot when finite bias voltage is applied
to the system.
It is pointed out that spin moment on the dot under finite bias
voltage does not necessarily produce spin polarized current.  We stress the
role of strong correlation for realizing it.
After identifying the proper characteristic temperature by universal
scaling in Sec.~\ref{sec:T-star}, which is imperative to identify and
understand the range of validity of our approximation scheme, we
present numerical results regarding spin transport by using the finite
interaction slave-boson approach in Sec.~\ref{sec:result}.  The
generation and suppression of spin-dependent transport are discussed.
Finally we conclude in Sec.~\ref{sec:conclusion}.

\section{Theory}
\label{sec:theory}

\begin{figure}
 \centering 
 \includegraphics[width=0.7\linewidth,clip]{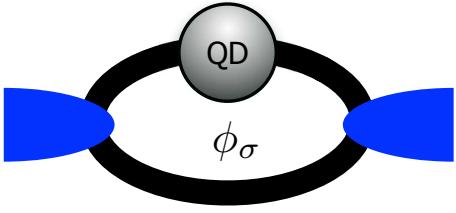}
  \caption{(Color online) Schematic illustration of the model. On the quantum dot
    (QD), Coulomb interaction is present and the Rashba spin-orbit
    interaction is incorporated as a spin-dependent flux
    $\phi_{\sigma}$ (see the text).}
\label{fig:QD-ABI}
\end{figure}

\subsection{Model}

We assume a single dot orbital to predominantly contribute to
transport near the Fermi level, and ignore the Zeeman splitting.
Accordingly, our theoretical model is the single-level Anderson model
augmented by the direct hopping between left and right leads ($\ell =
L,R$) and the Rashba SO interaction as $\phi_{\sigma}$ (see
Fig.~\ref{fig:QD-ABI} and below).
The total Hamiltonian of the system is composed of $H = H_{D} + H_{T}
+ \sum_{\ell} H_{\ell} + H_{A}$, where $H_{D}$ represents the dot
Hamiltonian; $H_{T}$, the hopping between the dot and the leads;
$H_{\ell}$, the noninteracting electron on the lead ($\ell=L,R$), and
$H_{A}$, for the arm with direct hopping between left and right leads.
We treat $H_{\ell}$ within the wide-band approximation (see
eq.~\eqref{eq:wide-band-limit-g}), and other terms are specified by
\begin{align}
& H_{D} = \sum_{\sigma} \epsilon_{d}\, n_{\sigma} + U n_{\uparrow}
n_{\downarrow},\\ 
& H_{T} =  \sum_{\ell,\sigma} \left(
    V_{d\sigma \ell}\, d^{\dagger}_{\sigma}  c_{\ell\sigma} + V_{\ell
      d\sigma}\, c^{\dagger}_{\ell\sigma} d_{\sigma} \right),\\
&   H_{A} = \sum_{\sigma} \left[ V_{LR}\, c^{\dagger}_{L\sigma}
    c_{R\sigma} + V_{LR}\, c^{\dagger}_{R\sigma}
    c_{L\sigma} \right],
\end{align}
where $n_{\sigma} = d^{\dagger}_{\sigma} d_{\sigma}$ is the dot
electron number operator.  Finite bias voltage is applied by
$\mu_{R,L} = \pm eV/2$ between the two chemical potentials of the
leads, and we control the gate voltage by shifting the dot level
$\epsilon_{d}$. The average dot number $n_{d} = \langle n_{\uparrow} +
n_{\downarrow}\rangle$ roughly corresponds to $2$, $1$, $0$ for
$\epsilon_{d} \lesssim -U$, $-U \lesssim \epsilon_{d} \lesssim 0$, and
$0 \lesssim \epsilon_{d}$ respectively.
We incorporate the local Rashba spin-orbit interaction
effectively as a spin-dependent phase $\phi_{\sigma}=\sigma
\phi_{\text{so}}$ (for $\sigma = \pm 1$) through the
interferometer~\cite{Sun05a}, where $\phi_{\text{so}}$ is proportional
to the asymmetric electric field perpendicular to the semiconductor
interface.  We adopt the convention
\begin{align}
 V_{Rd\sigma} V_{d\sigma L} V_{LR} = \big| V_{Rd} V_{dL}
 V_{LR} \big| \, e^{i\phi_{\sigma}},
\end{align}
where $|V_{\ell d\sigma}|=|V_{d\sigma \ell}|$ etc.\ are
spin-independent and we omit writing the suffix $\sigma$.

\subsection{Charge and spin currents}

We briefly summarize the formulae of charge and spin currents in terms
of nonequilibrium Green functions, which can be derived along the
standard line of treatment~\cite{Bulka01,Hofstetter01,Kim03}.
Here so as to clarify the role of $\phi_{\text{AB}}$ and
$\phi_{\text{so}}$, we proceed for a more general case of coexisting
$\phi_{\text{AB}}$ and $\phi_{\text{so}}$, that is, 
$\phi_{\sigma}= \phi_{\text{AB}}+\sigma \phi_{\text{so}}$ through the
interferometer.
The system is characterized by the following \emph{spin-independent}
parameters: 
\begin{align}
& \gamma = \gamma_{L} + \gamma_{R}; \qquad \gamma_{\ell} = \pi
|V_{d\ell}|^{2} \rho_{\ell},\\ 
& \xi = 4\pi^{2} \rho_{L} \rho_{R} |V_{RL}|^{2},
\end{align}
as well as asymmetry factor of the leads $\alpha = 4 \gamma_{R}
\gamma_{L}/\gamma^{2}$.  In the following, we use the wide-band limit
approximation extensively, assuming constant DOS and relaxation rate.

The current formula usually involves the lesser and retarded parts of
nonequilibrium Green function.  The present Hamiltonian, however,
conserves both the total charge and $z$-component of total spin
respectively, so we may eliminate the dependence of the lesser Green
function by using the conservation laws.
As a result, we can express the spin-resolved current $I_{\sigma} \equiv
I_{L\sigma} = -e \langle \dot{n}_{L\sigma}\rangle$ in terms of only
the retarded dot Green function. The formula reads
\begin{align}
& I_{\sigma} = -\frac{e}{h} \int d\varepsilon\,
  \left[ f_{L}(\varepsilon) -
  f_{R}(\varepsilon) \right] \mathcal{T}_{\sigma}(\varepsilon),
\label{eq:spin-current}
\end{align}
where $f_{\ell}(\varepsilon)$ is the Fermi distribution on the lead
$\ell$.  Transmission $\mathcal{T}_{\sigma}(\varepsilon) =
\mathcal{T}_{b} + \mathcal{T}_{1\sigma}(\varepsilon)$ consists of two
contributions: the background transmission $\mathcal{T}_{b} =
4\xi/(1+\xi)^{2}$ due to the arm, and $\mathcal{T}_{1\sigma}$ through
the dot, which is spin-dependent in general and found to be expressed
in terms of the \emph{exact} retarded Green function $G^{R}_{\sigma\sigma}$ of
a dot electron with spin $\sigma$ as
\begin{equation}
  \mathcal{T}_{1\sigma}(\varepsilon) = \mathcal{T}_{b}\, \Gamma\, \mathrm{Im} 
  \left[(1+iq_{\sigma}) (1+iq^{*}_{\sigma}) G^{R}_{\sigma\sigma}
    (\varepsilon) \right].
\label{eq:T-1sigma}
\end{equation}
Here $\Gamma=\gamma/(1+\xi)$ is the reduced relaxation rate in a ring
geometry, and the parameter and $q_{\sigma} = \sqrt{\alpha/\xi}
(e^{i\phi_{\sigma}} - \xi e^{-i\phi_{\sigma}})/2$ is the spin-dependent
Fano parameter.  Nonlinear spin-resolved conductance $\mathcal{G}_{\sigma}$ is
obtained by
\begin{align}
\mathcal{G}_{\sigma} = \frac{dI_{\sigma}}{dV} = \mathcal{G}_{b} +
\mathcal{G}_{1\sigma}, 
\label{def:G-sigma}
\end{align}
where $\mathcal{G}_{b} = (e^{2}/h) \mathcal{T}_{b}$ is the
background conductance due to the arm. Charge conductance $\mathcal{G} =
\sum_{\sigma} \mathcal{G}_{\sigma}$ oscillates with varying
$\phi_{\sigma}$, that is, $\phi_{\text{AB}}$ or $\phi_{\text{so}}$.
Such phenomena were experimentally observed~\cite{Konig06,Bergsten06}.

The above formula of spin-resolved current,
eq.~\eqref{eq:spin-current}, is exact; a correction such as a
two-particle Green function does not appears even if the strong
correlation is present on a dot. The situation is the same as the
Meir-Wingreen formula of the charge current~\cite{Meir92}, which we
reproduce in the limit of $\xi\to 0$ and $\mathcal{T}_{b} |q|^{2} \to
\alpha$.  The formula encodes full account of the strong correlation
effect in the exact one-particle retarded Green function
$G^{R}_{\sigma\sigma}(\varepsilon)$.
Another observation is that the presence of the interaction does not
affect the Fano parameter $q$; it is determined entirely by the
geometry of conducting leads.

Seeing the exact formula eq.~\eqref{eq:spin-current}, we get a basic
idea of how to generate spin transport electrically in a nonmagnetic
system ($\phi_{\sigma} = \sigma \phi_{\text{so}}$).  The factor
$(1+iq_{\sigma})(1+iq_{\sigma}^{*})$ depends only on
$\cos\phi_{\sigma}$, hence spin-independent. The formation of spin
moment on the dot does not immediately ensure the spin polarized
current.  We cannot anticipate any spin-dependent transport until the retarded
Green function $G^{R}_{\sigma\sigma}$ acquires spin-dependence.

\subsection{Finite $U$ slave-boson mean field approach}

In order to examine spin transport, 
our remaining task is to evaluate the exact one-particle retarded Green
function $G^{R}_{\sigma\sigma}(\varepsilon)$ appearing in
eq.~\eqref{eq:T-1sigma} in the
presence of the Coulomb interaction and finite bias voltage, which is
far from being trivial.
The two powerful methods successful in equilibrium systems, the Bethe
ansatz approach and numerical renormalization group calculations, have
some difficulty in applying to nonequilibrium systems with finite bias
voltage.  To the best of our knowledge, the exact evaluation has not
been so far available.  Accordingly we need to seek an appropriate
approximation scheme; we choose to adopt the Kotliar-Ruckenstein (KR)
formulation of the finite $U$ slave-boson mean field theory (SBMT),
originally introduced in equilibrium systems~\cite{Kotliar86} and
extended to nonequilibrium transport later.~\cite{Dong01}

The KR formulation of slave-boson approach enables us to evaluate the
Green function $G^{R}$ approximately, which leads to a Fermi-liquid
form (see eq.~\eqref{eq:qp-form-of-G} below), satisfying the
Friedel-Langreth sum rule.  It is known to reproduce correctly various
low-temperature behaviors including conductance enhancement due to the
Kondo effect.  It gives reliable results not only qualitatively but
also quantitatively, agreeing with linear conductance $\mathcal{G}$
obtained by numerical renormalization group
methods below the Kondo temperature~\cite{Dong01,Takahashi06,Oguchi10}.
The approach, retaining finite Coulomb interaction, can access the
full gate voltage dependence of conductance, which is important in
experiments.  One of the authors recently applied the approach to a
quantum dot with two-fold level degeneracy, successfully giving a
reasonably good account of linear and nonlinear conductance observed
in experiments in the entire range of gate
voltage~\cite{Oguchi10,Oguchi09}.

Following a standard treatment of the KR-SBMT, we introduce four
bose fields associated to each state of the dot: $e$ for the empty,
$p_{\sigma}$ for one electron with spin $\sigma$ and $d$ for the
doubly occupied state.  In the physical subspace, fermion operators
$d_{\sigma}$ and $d_{\sigma}^{\dagger}$ are replaced by quasiparticle
operator $f_{\sigma}z_{\sigma}$ and $z^{\dagger}_{\sigma}
f_{\sigma}^{\dagger}$ with the renormalization factor $z_{\sigma}$,
which is chosen to be~\cite{Kotliar86}
\begin{equation}
  z_{\sigma} = (1-d^{\dagger} d - p^{\dagger}_{\sigma}
  p_{\sigma})^{-\frac{1}{2}} (e^{\dagger} p_{\sigma} +
  p^{\dagger}_{\bar{\sigma}} d )  (1-e^{\dagger} e -
  p^{\dagger}_{\bar{\sigma}} p_{\bar{\sigma}})^{-\frac{1}{2}}. 
\end{equation}
In order to eliminate unphysical states, the completeness condition
$e^{\dagger}e + \sum_{\sigma}p^{\dagger}_{\sigma}p_{\sigma} +
d^{\dagger}d = I$, and the charge correspondence $f^{\dagger}_{\sigma}
f_{\sigma}=p^{\dagger}_{\sigma}p_{\sigma} + d^{\dagger} d$ must be
imposed by Lagrange multipliers $\lambda^{(1)}$ and
$\lambda^{(2)}_{\sigma}$.  
Accordingly, the Hamiltonian becomes, in terms of these slave boson
fields and quasiparticle operators, 
\begin{align}
& H_{D} \: \mapsto\:  \sum_{\sigma} \epsilon_{d}
f_{\sigma}^{\dagger} f_{\sigma} + U d^{\dagger} d, \\
& H_{T} \: \mapsto\: \sum_{\ell,\sigma} \left( \tilde{V}_{d\sigma\ell}
  f^{\dagger}_{\sigma} c_{\ell\sigma} + \tilde{V}_{\ell\sigma d}\,
  c^{\dagger}_{\ell\sigma} f_{\sigma} \right),
\end{align}
with the constraint Hamiltonian 
\begin{align}
& H_{\lambda} = \lambda^{(1)}\big( e^{\dagger} e + \sum_{\sigma}
    p^{\dagger}_{\sigma} p_{\sigma} + d^{\dagger}d -1 \big) 
\nonumber \\ & \quad
+  \sum_{\sigma} \lambda^{(2)}_{\sigma} \left( p_{\sigma}^{\dagger}
  p_{\sigma} + d^{\dagger}d - f^{\dagger}_{\sigma} f_{\sigma}
\right).  
\end{align}
Within the mean field approximation, all the boson fields are replaced
by their expectation values; then the Hamiltonian is reduced to the
renormalized resonant level model with the effective dot level
$\tilde{\varepsilon}_{d\sigma}=\epsilon_{d}-\lambda^{(2)}_{\sigma}$ as
well as the effective hopping $\tilde{V}_{d\sigma\ell} = z_{\sigma}
V_{d\ell}$.
The self-consistent equations to solve in conjugation with the
constraint are~\cite{Dong01,Takahashi06}
\begin{align}
& \tilde{\varepsilon}_{d\sigma} - \epsilon_{d} = \sum _{\sigma'}
\left( \frac{\partial 
  Z_{\sigma'}}{\partial |p_{\tau}|^{2}} - \frac{\partial
  Z_{\sigma'}}{\partial |e|^{2}} \right) \frac{\tilde{M}_{\sigma
}}{Z_{\sigma'}}, 
\label{eq:self-consistent1}    
\\ 
& U + \sum_{\sigma} \left(
    \frac{\partial Z_{\sigma}}{\partial |e|^{2}} - \sum_{\sigma'}
    \frac{\partial Z_{\sigma}}{\partial |p_{\sigma'}|^{2}} +
    \frac{\partial Z_{\sigma}}{\partial |d|^{2}}
  \right)\frac{\tilde{M}_{\sigma}}{Z_{\sigma}}  =0.
\label{eq:self-consistent2}    
\end{align}
where $Z_{\sigma}= |z_{\alpha}|^{2}$, and $\tilde{M}_{\sigma}$ is
defined by the quasi-particle Green function $\tilde{G}_{\sigma\sigma}$ as
\begin{align}
& \tilde{M}_{\sigma}
= \int^{\infty}_{-\infty}\frac{d\varepsilon}{2\pi i} (\varepsilon -
\tilde{\varepsilon}_{d\sigma} )
\tilde{G}^{<}_{\sigma \sigma}(\varepsilon).
\label{eq:M-tilde-sigma}
\end{align}
We determine numerically self-consistent bose fields $(e,
p_{\sigma}, d)$ satisfying eqs.~\eqref{eq:self-consistent1}
and~\eqref{eq:self-consistent2} at each temperature and bias voltage.

Once self-consistent boson fields are obtained, we see that the Green
function has a form of the renormalized Fermi liquid with the
quasiparticle weight $Z_{\sigma}$.  Namely, the renormalized
relaxation rate becomes $\tilde{\Gamma}_{\sigma} = Z_{\sigma} \Gamma$
and the renormalized energy, $\tilde{\varepsilon}_{d}$.  We recast
$G^{R}$ as
\begin{align}
& G^{R}_{\sigma\sigma}(\varepsilon) 
= \frac{1}{\varepsilon-\epsilon_{d} - \Sigma_{\sigma}(\varepsilon)}
\\ & \quad
 \approx \frac{Z_{\sigma}}{\varepsilon - \tilde{\varepsilon}_{d\sigma} +
    \tilde{\Gamma}_{\sigma} \sqrt{\alpha\xi} \cos\phi_{\sigma} + i
    \tilde{\Gamma}_{\sigma}}. 
\label{eq:qp-form-of-G}
\end{align}
When we plug the above form in eq.~\eqref{eq:T-1sigma}, we see
$\mathcal{T}_{\sigma}$ take a form of the Fano
formula~\cite{Fano61,Hofstetter01,Takahashi06}, 

\begin{equation}
  \mathcal{T}_{\sigma}(\varepsilon) = \mathcal{T}_{b} \,
  \frac{|e_{\sigma}+q_{\sigma}|^{2}}{e^{2}_{\sigma}+1}.
\end{equation}
Note $q_{\sigma}$ is the (spin-dependent) Fano parameter previously
defined, and $e_{\sigma} = (\varepsilon -
\tilde{\varepsilon}_{d\sigma})/\tilde{\Gamma}_{\sigma} +
\sqrt{\alpha\xi} \cos\phi_{\sigma}$ is a dimensionless detuning from
the resonance energy, which includes interaction effect.
It is stressed that interaction affects the detuning
$e_{\sigma}$ but not the Fano parameter $q_{\sigma}$.

\section{Spin Moment Induced by $\phi_{\sigma}$ and Its Transport}
\label{sec:spin-moment}

As a prerequisite to realizing spin transport, we now examine how the
local Rashba SO phase induces finite spin moment on the dot once
finite bias voltage is applies in nonmagnetic systems despite apparent
time-reversal symmetry.  We might often expect transport to become
spin-dependent once finite moment emerges because the latter normally
is associated with spin-dependent shift and/or relaxation.  However,
as we see immediately in \S\ref{sec:noninteracting-dot}, a
noninteracting SOI system turns out exception to this rule; finite
spin moment on a noninteracting SOI arises no spin transport.
Accordingly we need some additional effect --- correlation effect ---
to realize it.
In this section, we will clarify the effect by deriving an effective
Keldysh action of the dot under finite bias.

\subsection{Effective Keldysh action of the dot under finite bias}
\label{sec:noninteracting-dot}

Since conducting electrons on the leads are assumed to be
noninteracting, it is straightforward to integrate exactly over those degrees
of freedom.  The resulting effective action describes the
nonequilibrium formation of spin polarization on an open
quantum dot.  By assuming the spin dependent phase $\phi_{\sigma} =
\phi_{\text{AB}}+ \sigma \phi_{\text{so}}$ for the sake of generality,
integrating over the lead degrees of freedom produces the effective
Keldysh action of the dot
\begin{align}
 S_{\text{eff}} &= \int_{c} dt dt' \sum_{\sigma}\bar\Psi_{\sigma}(t)
\left( \hat{G}_{\text{iso}}^{-1} 
- \hat{\Sigma}_{0} \right) (t,t') \Psi_{\sigma}(t') 
\nonumber \\ & \quad
 - U \int_{c} dt\, \bar\Psi_{\uparrow}(t) \bar\Psi_{\downarrow}(t)
 \Psi_{\downarrow}(t) \Psi_{\uparrow}(t).
\label{eq:effective-S}
\end{align}
Here $\Psi$ is a spinor field associated with the dot electron of the
Keldysh doublet structure; $\hat{G}_{\text{iso}}$ is the $2\times 2$ Green function
matrix describing a noninteracting, isolated dot.  The self-energy
$\hat{\Sigma}_{0}$ is an outcome of integrating the lead degrees of freedom,
which is found to become in terms of Green functions of a decoupled lead
$g_{\ell}$:
\begin{align}
& \hat{\Sigma}_{0} = \sum_{\ell,\ell'=L,R} \hat{V}_{d\ell} \,
\hat{\bar{g}}_{\ell\ell'} \,
\hat{V}_{\ell' d}, \\
& \hat{\bar{g}}^{-1} = 
\begin{pmatrix} 
  g^{-1}_{L} & - \hat{V}_{LR}\otimes \tau_{3} \\ 
  - \hat{V}_{RL} \otimes \tau_{3} &  g^{-1}_{R} \end{pmatrix},
\end{align}
where $\tau_{i}$ are the Pauli matrices in the Keldysh structure. In
the wide-band limit, $g$ approximates to
\begin{equation}
 \hat{g}_{\ell} \approx 
-i\pi\rho_{\ell} 
\label{eq:wide-band-limit-g}
\begin{pmatrix}
  1-2f_{\ell} & 2f_{\ell} \\ 2(1-f_{\ell}) & 1-2f_{\ell}
\end{pmatrix},
\end{equation}
so that one finds the retarded ($R$), advanced ($A$), Keldysh ($K$)
parts of the self-energy $\Sigma_{0}$ to be
\begin{align}
&  \Sigma^{R,A}_{0,\sigma} = -\Gamma \sqrt{\alpha\xi} \cos\phi_{\sigma} \mp
i\Gamma, 
\label{eq:Sigma0-R}\\
& \Sigma^{K}_{0,\sigma} = -2i \Gamma (1-2\bar f_{\sigma}).
\end{align}
Here, we have introduced the effective dot spin distribution
$\bar{f}_{\sigma}$~\cite{footnote1}
by 
\begin{align}
& \bar  f_{\sigma}(\phi_{\sigma}) =
\frac{\Sigma^{<}_{0,\sigma}}{\Sigma^{A}_{0,\sigma} - \Sigma^{R}_{0,\sigma}} 
= \frac{1}{\Gamma} \sum_{\ell} 
f_{\ell} \, \Gamma'_{\ell\sigma}(\phi_{\sigma}),
\label{eq:f-sigma}
\end{align}
and relaxation rates 
\begin{align}
& \Gamma  = \sum_{\ell} \Gamma'_{\ell\sigma} = \frac{\gamma}{1+\xi},
\label{def:Gamma}\\
& \Gamma'_{L\sigma} (\phi_{\sigma})
= \frac{\gamma_{L} + \xi \gamma_{R} - \gamma \sqrt{\alpha \xi
 } \sin\phi_{\sigma}}{(1+\xi)^{2}}, \\
& \Gamma'_{R\sigma} (\phi_{\sigma})
= \frac{\gamma_{R} + \xi \gamma_{L} + \gamma \sqrt{\alpha \xi
 } \sin\phi_{\sigma}}{(1+\xi)^{2}}.
\end{align}
We note that the relaxation rate $\Gamma$ does not depend on spin
$\sigma$.
Accordingly $\Sigma_{0}$ in eq.~\eqref{eq:effective-S} is equal to
\begin{align}
& \hat{\Sigma}_{0} = \frac{1}{2}
\begin{pmatrix} 1 & 1 \\ -1 & 1 \end{pmatrix}
\begin{pmatrix}
  \Sigma^{K}_{0} & \Sigma^{R}_{0} \\ \Sigma^{A}_{0} & 0
\end{pmatrix} 
\begin{pmatrix} 1 & -1 \\ 1 & 1 \end{pmatrix}.
\end{align}

\subsection{Noninteracting dot}

When $U=0$, the effective action $S_{\text{eff}}$ with
$\hat{\Sigma}_{0}$ provides a full solution of the problem. In this
case, all of the nonequilibrium Green functions can be evaluated
exactly by $G_{0}=(G^{-1}_{\text{iso}}-\Sigma_{0})^{-1}$, to
find~\cite{Crisan09}
\begin{align}
& G^{<}_{0,\sigma\sigma}(\varepsilon) = -2i
\bar{f}_{\sigma}(\varepsilon)\, \Im  
G^{R}_{0,\sigma\sigma}(\varepsilon), 
\label{eq:Gless-GR}\\
&  G^{R}_{0,\sigma\sigma}(\varepsilon) =
  \frac{1}{\varepsilon-\varepsilon_{d} + \Gamma\sqrt{\alpha\xi}
      \cos\phi_{\sigma} + i\Gamma}.
\label{eq:GR-U0}
\end{align}
Spin part of the distribution $s_{0}(\varepsilon) =
\bar{f}_{\uparrow}(\varepsilon) - \bar{f}_{\downarrow}(\varepsilon)$ is
\begin{equation}
s_{0}(\varepsilon) 
= \frac{2\sqrt{\alpha\xi}}{1+\xi} \left[ f_{R}(\varepsilon) -
  f_{L}(\varepsilon) \right] \cos\phi_{\text{AB}}
\sin\phi_{\text{so}}.
\label{eq:spin-density}
\end{equation}

It is clear that even after switching off $\phi_{\text{AB}}$ having
the system nonmagnetic, finite spin polarization remains~\cite{Crisan09}.
Spin polarization $m_{0}=\langle n_{\uparrow} - n_{\downarrow}
\rangle$ in this case becomes
\begin{align}
&m_{0} 
= \frac{2\sqrt{\alpha\xi}\sin\phi_{\text{so}}}{\pi(1+\xi)} \Im \log \left[
\frac{\mu_{L} - \varepsilon_{d}-
  \Sigma^{R}_{0}}{\mu_{R}-\varepsilon_{d}-\Sigma^{R}_{0}} \right].
\end{align}
We come to the conclusion that finite spin polarization appears on a
noninteracting dot when the following three conditions are met: (1) in
a ring geometry $\xi \neq 0$, (2) under finite bias voltage
$f_{R}-f_{L} \neq 0$, and (3) with finite Rashba SO coupling,
$\phi_{\text{so}} \neq 0$. 

By contrast, switching off $\phi_{\text{AB}}$ renders transport
spin-independent in spite of the above finite spin polarization.  This
is seen because $G^{R}_{\sigma\sigma}(\varepsilon)$ loses all the
spin-dependence in this case (see
eqs.~(\ref{eq:spin-current},\ref{eq:T-1sigma}) and the Fano parameter
$q_{\sigma}$ is spin-independent), leading to $I_{\uparrow} =
I_{\downarrow}$ .
Finite spin polarization with vanishing spin current is a peculiar
feature of a noninteracting SOI system.

\subsection{Nonmagnetic interacting dot}

When one turns on interaction $U$ on the dot in nonmagnetic systems
($\phi_{\text{AB}}=0$), one may expect spin-dependent transport 
due to finite spin polarization.  This is because finite spin
polarization affects the system through interaction channel 
inducing spin-dependent shift and relaxation process in the
exact one-particle retarded function $G^{R}_{\sigma\sigma}(\varepsilon)$. 

One can confirm the above statement by the lowest-order perturbation
calculation regarding interaction, and by the self-consistent
extension of it.  Such perturbational result can be justified at the
temperature higher than the Kondo temperature, though.
When one writes the exact Green function symbolically as $\hat{G} = (
\hat{G}^{-1}_{0} - \hat{\Sigma}_{0}-\hat{\Sigma}_{U} )^{-1}$ in the
Keldysh space, the Hartree contribution leads to $\Sigma_{U,\sigma} =
U\langle n_{\bar{\sigma}} \rangle_{0}$ where $\langle \cdots
\rangle_{0}$ is the noninteracting average.  Up to this order, spin
current $I_{s}=I_{\uparrow} - I_{\downarrow}$ is found to be
\begin{align}
& I_{s} = I_{\uparrow} - I_{\downarrow} = 
-\frac{e}{h} \int d\varepsilon\, [f_{L}(\varepsilon) -
f_{R}(\varepsilon)]\, \Delta \mathcal{T}(\varepsilon),\\
& \Delta \mathcal{T}(\varepsilon) \approx  -\mathcal{T}_{b}
\frac{ m_{0} U}{\Gamma} \Im 
\left[ (1+iq)(1+iq^{*}) (\Gamma G^{R}_{0}(\varepsilon))^{2} \right],
\end{align}
where $G^{R}_{0}$ is the noninteracting Green function of
eq.~\eqref{eq:GR-U0} with $\phi_{\text{AB}}=0$; $m_{0}$, the spin polarization 
of eq.~\eqref{eq:GR-U0}.  Spin current $I_{s}$ behaves as
$m_{0} U(eV)^{2}/\Gamma^{3}$ for small bias voltage, so that though
the effect may be small, it should be counted as truly a
``nonequilibrium correlation effect.''  One may further calculate spin
polarization $m$ self-consistently, to find finite spin polarization
is stable~\cite{Crisan09}.  Hence finite spin current with
$m_{0}$ being replaced by $m$.


We argue that spin polarization remains finite at the temperature much
lower than the Kondo temperature, where the system enters the
strong-coupling regime.  
Indeed the Fermi liquid quasiparticle picture, which is effective for
the lower temperature, justifies the form eq.~\eqref{eq:spin-density}
of the effective distribution as a reasonably good approximation, if
not exact. The argument is as follows.
When we incorporate the self-energy $\Sigma_{U}(\varepsilon)$ up to
linear $\varepsilon$ in terms of the
renormalization factor $Z_{\sigma} = [1-\partial_{\varepsilon}
\Sigma_{U,\sigma}(\varepsilon)]^{-1}$, we can write the action as
(with suppressing spin indices)
\begin{align}
\bar{\Psi}\left( G_{0}^{-1} - \Sigma_{0} -
    \Sigma_{U} \right) \Psi
\approx \bar{\Psi}\, Z^{-1} 
  \tilde{G}^{-1} \Psi,
\end{align}
where we have introduced the quasiparticle Green function $\tilde{G} =
G/Z=(G^{-1}_{0} - \Sigma_{0} - \Sigma_{U})^{-1}/Z$ and the
renormalized self-energy $\tilde{\Sigma} = Z\, \Sigma_{0}$ on the
right hand side (with suppressing spin indices).  In this
quasiparticle picture, effective distribution
$\bar{f}_{\sigma} = \tilde{\Sigma}^{<}_{\sigma} /
(\tilde{\Sigma}^{A}_{\sigma}-\tilde{\Sigma}^{R}_{\sigma})$   
is well approximated by its bare one, since the renormalization factor
$Z_{\sigma}$ is cancelled out between the numerator and the
denominator.
\begin{align}
  \bar{f}_{\sigma}(\varepsilon;U) \approx
    \frac{\Sigma^{<}_{0,\sigma}}{\Sigma^{A}_{0,\sigma} -
      \Sigma^{R}_{0,\sigma}} 
= \bar{f}_{\sigma}(\varepsilon;U=0),
\end{align}
%
It means that finite spin polarization is persistent at very low
temperature.
This type of approximation is often called the Ng's ansatz and is
often utilized in evaluating the lesser Green function $G^{<}$ in the
presence of interaction~\cite{Ng93,Dinu07}.

Having established that finite spin polarization is present on the
interacting SOI system both in the higher and the lower temperature
regimes, we now argue that large spin transport is expected at $T=0$
within the Kondo valley region $\varepsilon_{d} \approx -U/2$.  This
is because strong correlation is known to enhance conductance
as well as provide spin-dependent shift and relaxation due to finite
moment on the dot that the Rashba SO interaction induces. 
Apparently, some aspect is similar to a large spin filtering effect of
linear conductance proposed in the Anderson system in a magnetic
field~\cite{Costi01}.  Once finite spin moment is present whether due
to magnetic field or the bias voltage, large spin dependence appears
in transport.  The difference is that in nonmagnetic SOI systems, we
do need finite bias to sustain spin moment, so that such spin
transport only appears as a nonlinear response.

\section{Characteristic Temperature $T^{*}(\phi)$ and Universal Scaling}
\label{sec:T-star}



In order to explore spin-dependent phenomena of the interferometer at
$T=0$ or very low temperature and to show an unambiguous connection to
the Kondo physics, we first need to identify the characteristic
temperature of the system, separating the weak-coupling and
strong-coupling regions. This also enables us to correctly identify
the range of validity of the slave-boson approach. The scale 
usually corresponds to the Kondo temperature.  The characterization, however,
is not so transparent particularly in a ring geometry.  Firstly, the
Kondo temperature characterizes a crossover, not a transition, so that
one can only determine the scale modulo numerical factor.  Secondly,
the flux phase either $\phi_{\text{AB}}$ and/or $\phi_{\text{so}}$
through the interferometer is known to affect the Kondo temperature in
a substantial and nontrivial way.

For this purpose, we will use the scale $T^{*}$, defined by eq.~\eqref{def:T-star}
below, which is endorsed by confirming universal scaling of
conductance.
It is known that conductance through a single quantum dot exhibits the
universal dependence on temperature or on bias voltage, once scaled by
the Kondo temperature~\cite{Costi94,Goldhaber-Gordon98b,Rincon09}.
By putting the other way round, observed universal dependence justifies
the choice of the characteristic temperature.
Recalling low-temperature physics is dominated by the Fermi liquid
picture, we introduce the characteristic temperature $T^{*}$ at
each gate voltage by the inverse scale of quasiparticle Green function
$\tilde{G} = G/Z$ at zero temperature in
equilibrium~\cite{footnote2}, which according to
eq.~\eqref{eq:qp-form-of-G} leads to the
phase-dependent characteristic temperature
\begin{equation}
  T^{*}(\phi) =  \left|-\tilde{\varepsilon}_{d}
    +\tilde{\Gamma} \sqrt{\alpha\xi} \cos\phi +
    i\tilde{\Gamma} \right|_{T=0, V=0}.
\label{def:T-star}
\end{equation}
When $\phi_{\text{AB}}$ and $\phi_{\text{so}}$ coexist, there are two
scales $T(\phi_{\sigma})$ for $\sigma = \uparrow, \downarrow$.  If only either $\phi_{\text{AB}}$ or $\phi_{\text{so}}$ is
present (this is the situation we are concerned with), we have
only one scale of the characteristic temperature.  
In the single impurity Anderson model (SIAM), the above definition
reduces to the Kondo peak width, $T^{*} = \tilde{\Gamma}$.

%
\begin{figure}
  \centering
\includegraphics[width=0.8\linewidth]{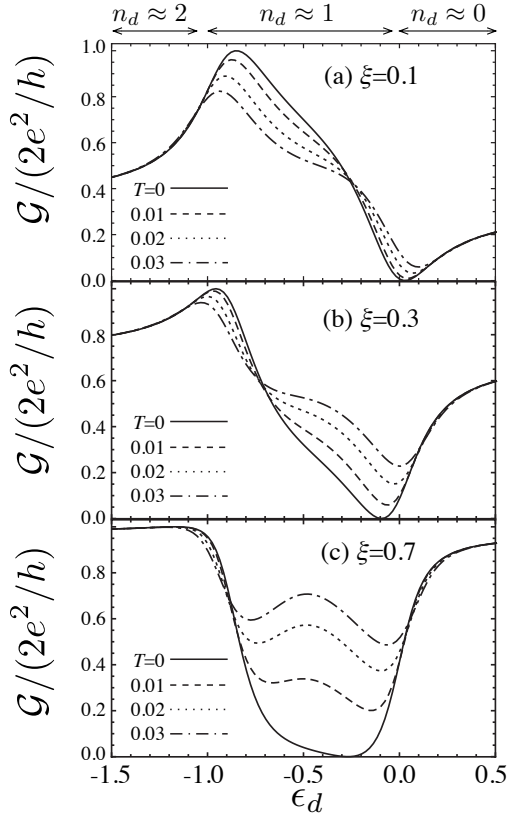}
\caption{Temperature evolution of linear conductance profile at
  $\phi=0$, for (a) $\xi=0.1$, (b) $\xi=0.3$, and (c) $\xi=0.7$.
  Energies are measured in unit of $U$.}
\label{fig:G-Ed}
\end{figure}
Figure~\ref{fig:G-Ed} shows numerical results of typical behavior of
linear conductance $\mathcal{G}=\mathcal{G}_{\uparrow} +
\mathcal{G}_{\downarrow}$ by using the KR-SBMT approach.  Depending on
a choice of $\xi$, conductance is enhanced at $\xi=0.1$ (the Kondo
effect), or suppressed (the anti-Kondo effect) at $\xi=0.3$ and $0.7$
with decreasing temperature.
%
 \begin{figure}
 \includegraphics[width=0.8\linewidth]{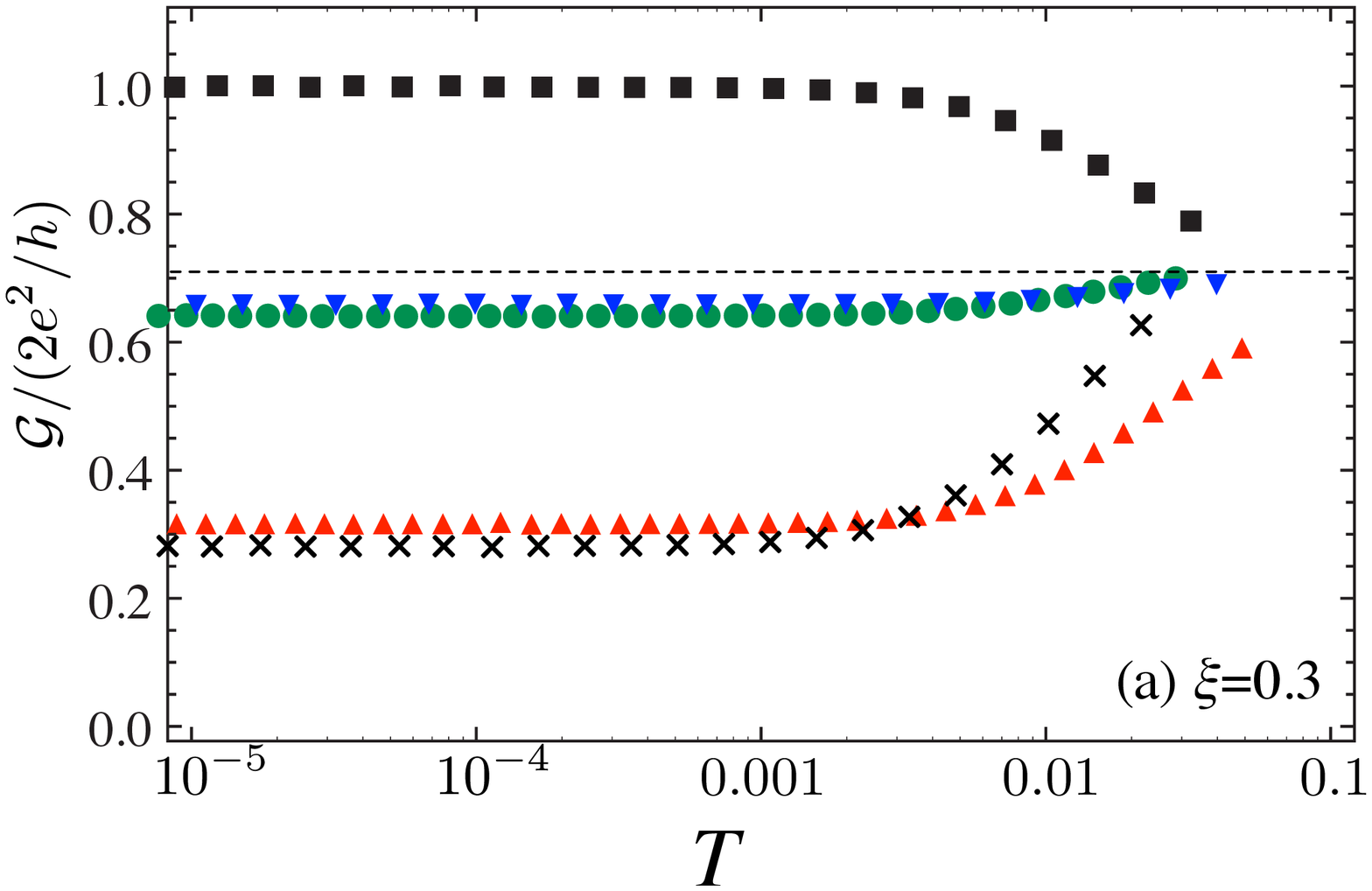}
 \includegraphics[width=0.8\linewidth]{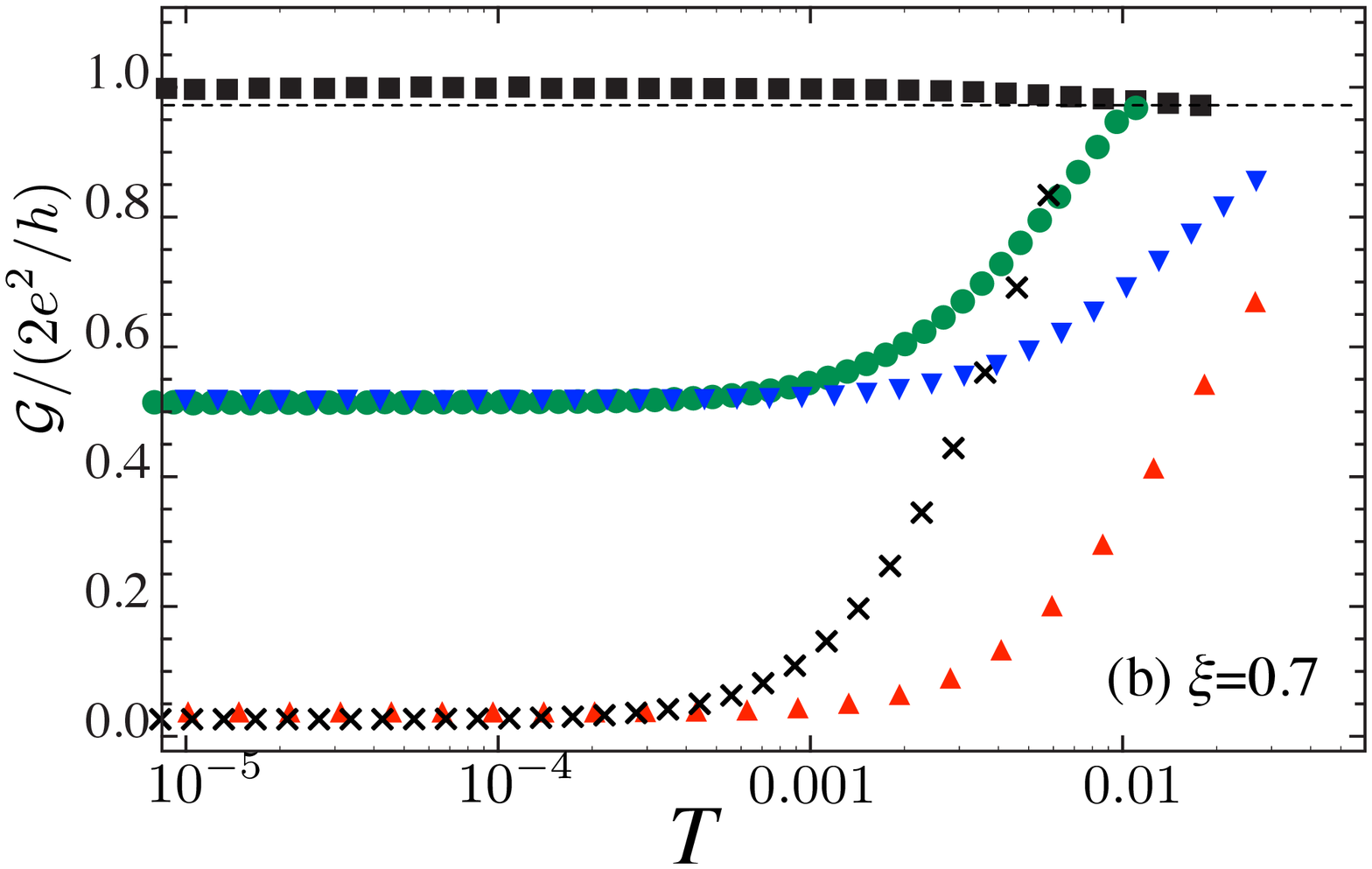}
 \caption{(Color online) Temperature dependence of linear conductance
   at $\epsilon_{d} = -U/2$, for (a) $\xi=0.3$ and (b) $\xi=0.7$.
   Behaviors are plotted at values $\phi=0$~($\triangle$),
   $\pi/4$~($\triangledown$), $\pi/2$~($\square$),
   $3\pi/4$~($\bigcirc$), and $\pi$~($\times$). Dashed lines
   correspond to $\mathcal{G}_{b}$.  }
 \label{fig:G-T-unscaled}
 \end{figure}
 In Fig.~\ref{fig:G-T-unscaled}, we take a closer look at the
 temperature dependence at the center of the Kondo valley
 $\epsilon=-U/2$, by varying $\phi=0$, $\pi/4$, $\pi/2$, and $3\pi/4$.
 Although we see linear conductance increase or decrease with
 increasing temperature, it clearly
 tends to approach the value corresponding to $\mathcal{G}_{b}$.
 Hence one expects such crossover behavior is dominated by the
 characteristic temperature of the system, and that when scaled by it,
 it exhibits a universal dependence.

%
 \begin{figure}
\begin{center}
\includegraphics[width=0.8\linewidth]{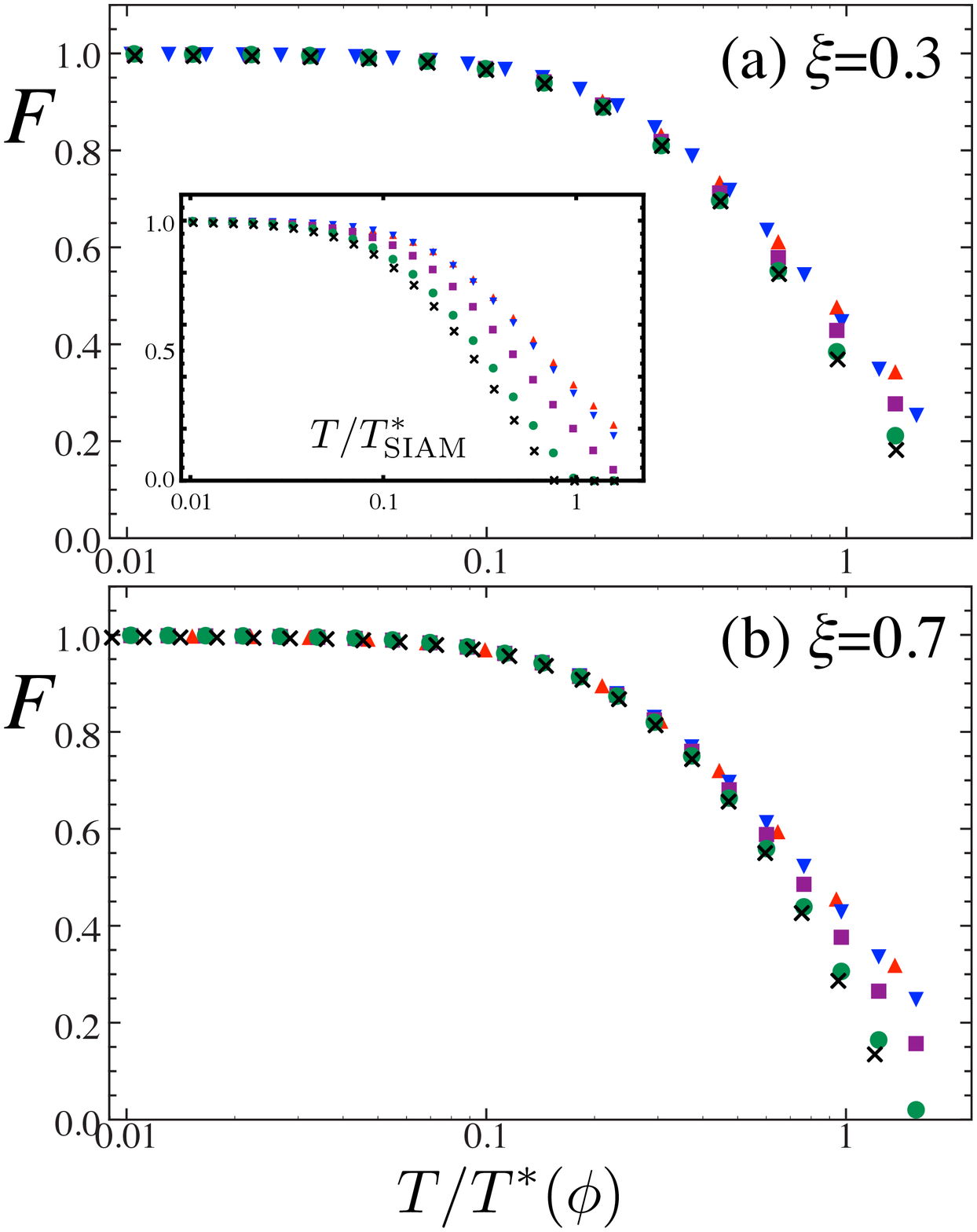}
\caption{(Color online) Universal temperature dependence of linear
  conductance for (a) $\xi=0.3$, and (b) $\xi=0.7$, scaled by
  $T^{*}(\phi)$. Behaviors are plotted at values
  $\phi=0$~($\triangle$), $\pi/4$~($\triangledown$),
  $\pi/2$~($\square$), $3\pi/4$~($\bigcirc$), and $\pi$~($\times$).
  Inset: No universal dependence is observed if one normalizes
  temperature by the characteristic temperature of SIAM
  $T^{*}_{\text{SIAM}}$.}
\label{fig:F-t}
\end{center}
\end{figure} 

We support this assertion by examining the scaling function 
\begin{equation}
  F(t=T/T^{*}(\phi)) = \left| \frac{\mathcal{G}(T,\phi) -2
      \mathcal{G}_{b}}{\mathcal{G}(0,\phi) - 2\mathcal{G}_{b}} \right|.
\end{equation}
(Factor $2$ in front of $\mathcal{G}_{b}$ comes from the spin degeneracy.) As shown in Fig.~\ref{fig:F-t}, they exhibit universal temperature
dependence for five different values of $\phi$.  It is noted that if one
alternatively scaled temperature by the characteristic temperature of
SIAM $T^{*}_{\text{SIAM}}$, one could not attain such universal behavior
(see the inset of Fig.~\ref{fig:F-t}~(a)).

We regard $T^{*}(\phi)$ as the proper scale of how the Kondo physics
takes effect in the present interferometer system.  By utilizing this
scale, we will see how spin-dependent transport is related to the
Kondo physics.

\section{Spin-dependent Transport}
\label{sec:result}


We are now going to examine how spin-dependent transport appears
electrically as a nonequilibrium effect by the Rashba coupling phase
$\phi_{\sigma} = \sigma \phi_{\text{so}}$ in nonmagnetic systems
$\phi_{\text{AB}}=0$.  We evaluate spin-resolved conductance
$\mathcal{G}_{\sigma} = dI_{\sigma}/dV$ and $\Delta \mathcal{G} =
\mathcal{G}_{\uparrow} - \mathcal{G}_{\downarrow}$ numerically within
the KR-SBMT approach according to eq.~\eqref{eq:spin-current}.  In the
following, we focus on a moderately strong Coulomb interaction case
$U/\gamma = 2.0$, and we measure energy in unit of $U=1$.  We will
discuss our results in light of the characteristic temperature
$T^{*}(\phi_{\text{so}})$ introduced and examined in the previous section.

\subsection{Finite bias voltage effect}

\begin{figure}
  \begin{center}
 \includegraphics[width=0.85\linewidth,clip]{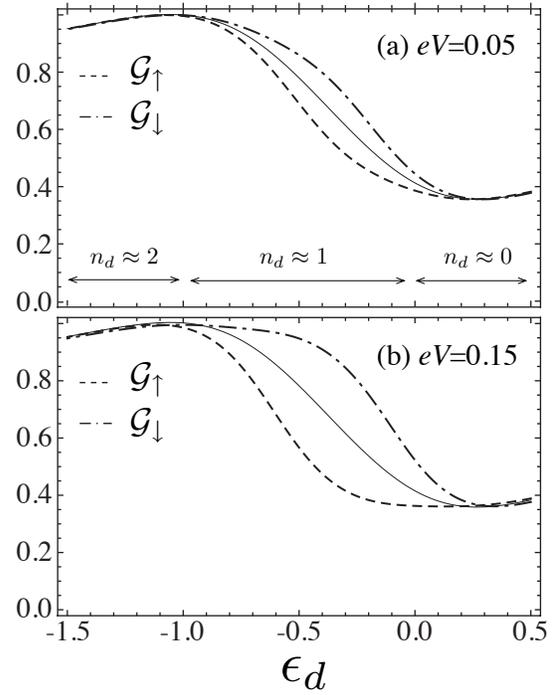}
  \end{center}
  \caption{Spin-resolved conductance $\mathcal{G}_{\sigma}$ (in unit
    of $e^{2}/h$) at zero temperature (dashed lines for
    $\mathcal{G}_{\uparrow}$; dashed-dotted lines for
    $\mathcal{G}_{\downarrow}$) as a function of the gate voltage
    $\epsilon_{d}$ for (top) $eV=0.05U$ and (bottom) $eV =
    0.15U$. Thin solid lines refer to linear conductance, where one
    see no spin-dependence. $\xi=0.3$ is chosen.}
\label{fig:GsvsEd-eV}
\end{figure}

Figure~\ref{fig:GsvsEd-eV} demonstrates how finite bias voltage
induces spin-dependence conductance $\mathcal{G}_{\sigma}$ at zero
temperature, with the choice of the Rashba SO phase $\phi_{\text{so}}
= \pi/4$.  While there is no spin-dependence in linear conductance
(shown by thin solid lines in Fig.~\ref{fig:GsvsEd-eV} (a) and (b)),
we see applying finite voltage gradually develop the spin-dependent
conductance in (a) $eV=0.05U$, and (b) $eV=0.15U$.
One further observes that such spin-dependent transport is conspicuous only
in the singly-occupied region of the dot ($n_{d} \approx 1$),
where the Kondo physics takes effect. 

\begin{figure}
  \centering
\includegraphics[width=0.8\linewidth,clip]{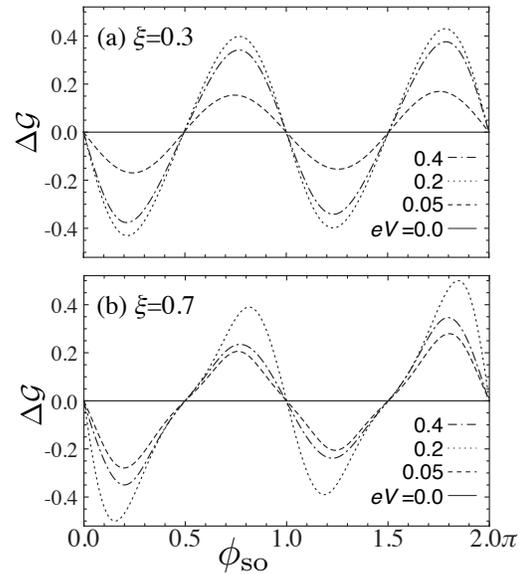}
\caption{The Rashba phase dependence of spin conductance $\Delta
  \mathcal{G}=\mathcal{G}_{\uparrow} - \mathcal{G}_{\downarrow}$ (in unit
    of $e^{2}/h$) for
  (a) $\xi = 0.3$ and (b) $\xi=0.7$. Different values of bias voltage
  are plotted by $eV=0$ (solid line),
  $0.05$ (dashed line), $0.2$ (dotted line), and $0.4$ (dash-dotted line), }
\label{fig:DG-SOI}
\end{figure}

We next show in Fig.~\ref{fig:DG-SOI} the Rashba SO phase dependence of
spin conductance $\Delta \mathcal{G}= \mathcal{G}_{\uparrow} -
\mathcal{G}_{\downarrow}$ at the middle of the Kondo valley
$\epsilon_{d} = -U/2$.  The results for $\xi=0.3$ and $0.7$ are
presented.  One sees $\Delta G$ oscillates as a function of
$\phi_{\text{so}}$. It vanishes not only at $\phi_{\text{so}} = 0$ and
$\pi$ when no spin density is on the dot (see
eq.~\eqref{eq:spin-density}), but also around $\pi/2$ and $3\pi/2$,
where the shift of the renormalized level vanishes altogether.
Furthermore, a closer look reveals that applying the larger bias
\emph{does not} necessarily produce the larger spin dependence.  Indeed,
the magnitude of the spin conductance for $V=0.4$ (dash-dotted line) turns
out smaller than that for $V=0.2$ (dotted line) over the entire value
of $\phi_{\text{so}}$ at $\xi=0.3$ and $0.7$.  It shows that there is
an optimal value to attain maximal spin-dependence.

\subsection{Temperature dependence}

\begin{figure}
\centering
 \includegraphics[width=0.8\linewidth,clip]{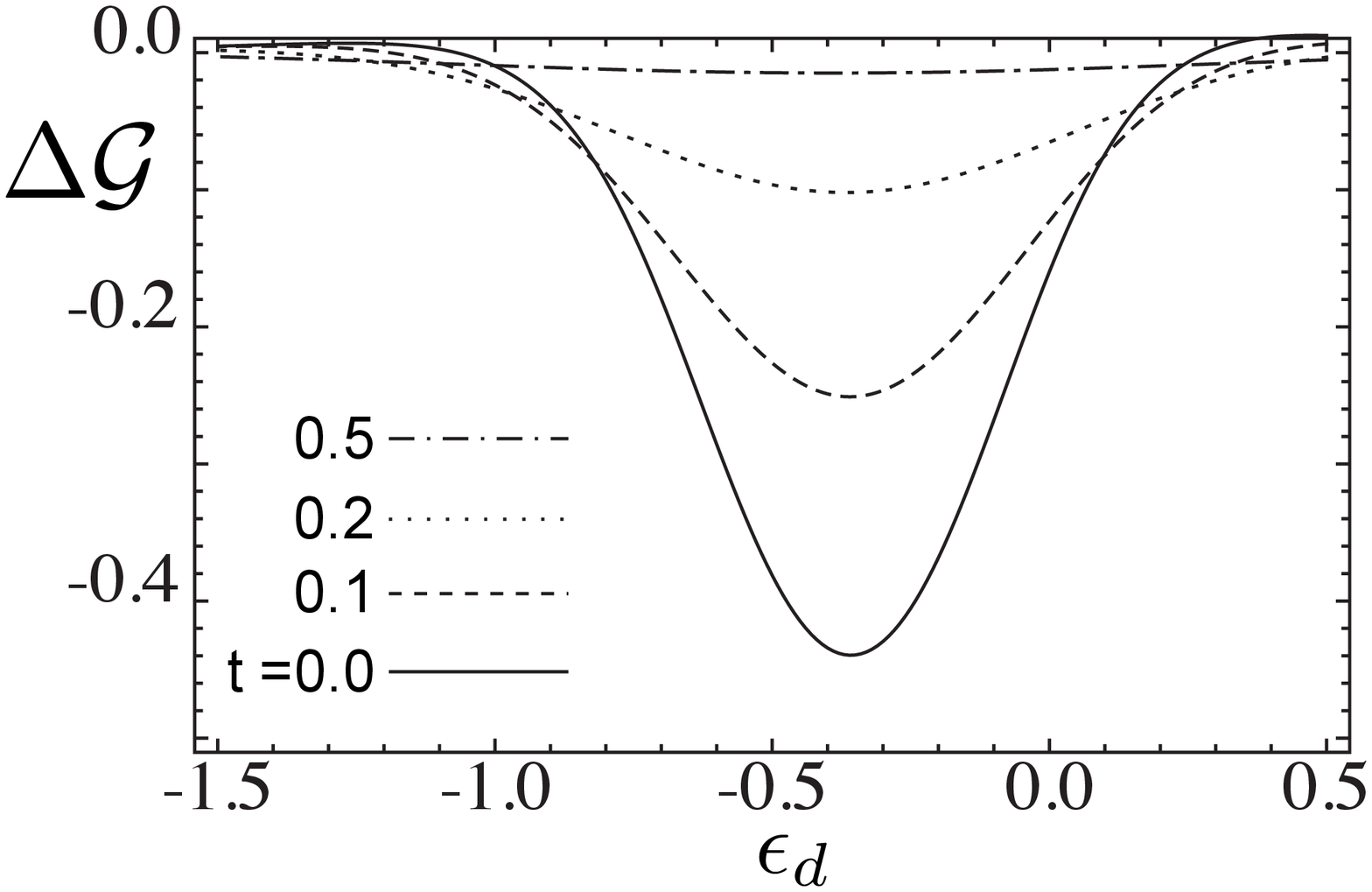}
 \caption{Temperature evolution of spin conductance profile $\Delta
   \mathcal{G} = \mathcal{G}_{\uparrow} - \mathcal{G}_{\downarrow}$ as
   a function of gate voltage $\epsilon_{d}$.  Temperature is
   scaled by $t=T/T^{*}(\phi_{\text{so}})$.  Other parameters are
   the same as in Fig.~\ref{fig:GsvsEd-eV} (b).}
\label{fig:dGvsEd-T}
\end{figure}

To clarify further the nature of the observed spin-dependent transport, we
examine its temperature evolution.  Figure~\ref{fig:dGvsEd-T}
represents spin conductance $\Delta \mathcal{G}$ as a function
of gate voltage by varying temperature.  Here, to clarify a connection
with the Kondo physics, we scale temperature by the characteristic
temperature $T^{*}(\phi_{\text{so}})$.
One observes a few things immediately: spin-dependence is maximal
around $\epsilon_{d} = -U/2$; Spin-dependence is eminent at
temperature lower than $T^{*}(\phi_{\text{so}})$, but it gets reduced
considerably toward approaching $T^{*}(\phi_{\text{so}})$.  These
observations strongly suggest that the Kondo physics is responsible
for the emergent spin transport.

\subsection{Spin-dependent transport as a nonequilibrium Kondo effect}

\begin{figure}
\centering
 \includegraphics[width=0.8\linewidth,clip]{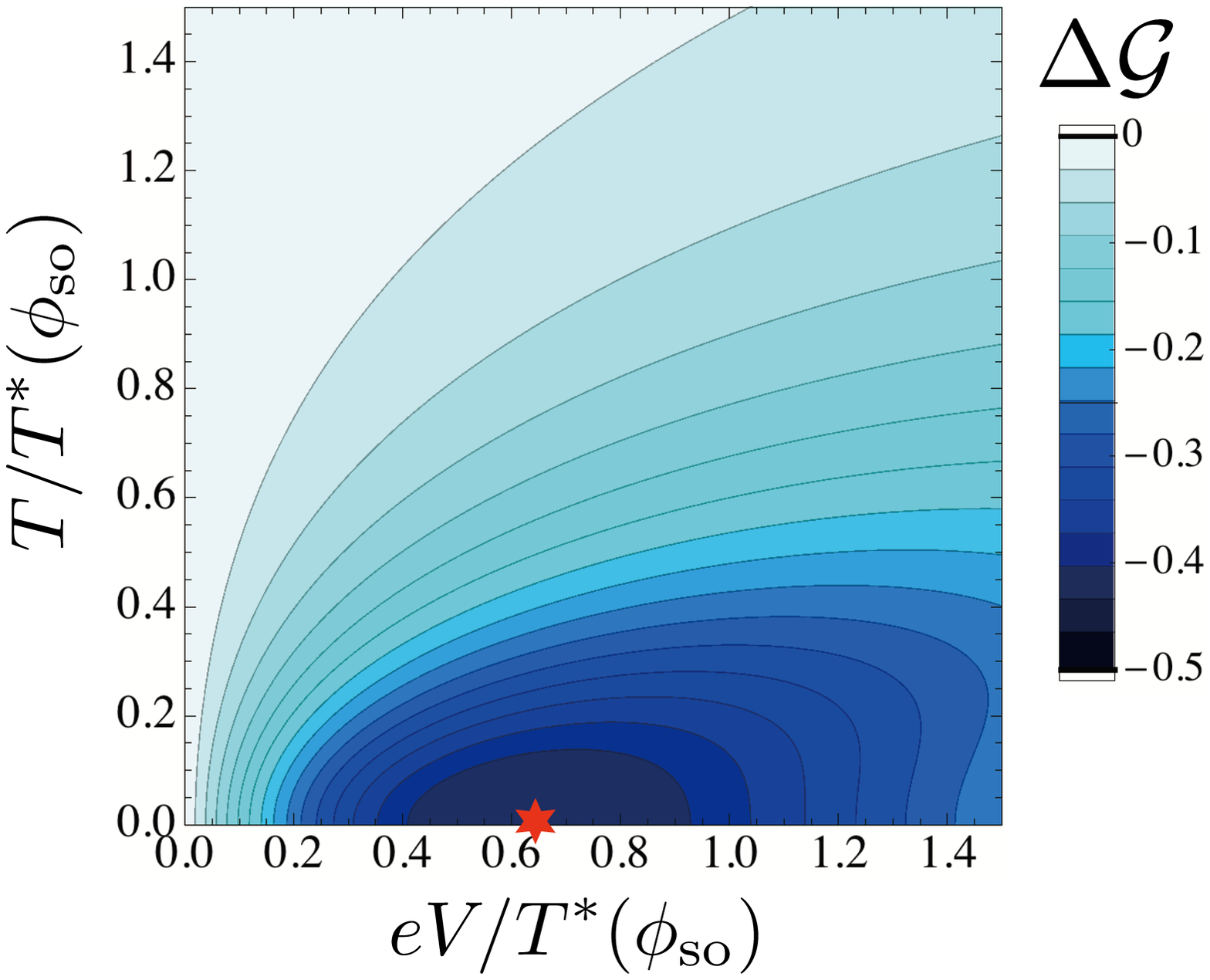}
\quad
\caption{(Color online) Dependence of $\Delta \mathcal{G} =
  \mathcal{G}_{\uparrow} - \mathcal{G}_{\downarrow}$ (in unit of
  $e^{2}/h$) by varying bias voltage and temperature.  Other
  parameters are the same as in Figs.~\ref{fig:GsvsEd-eV} (b)
  and~\ref{fig:dGvsEd-T}.  Both axes are normalized by the
  characteristic temperature $T^{*}(\phi_{\text{so}})$. Symbol $*$
  refers to the location at maximal amplitude of $\Delta
  \mathcal{G}$. }
\label{fig:dG-v-t}
\end{figure}

Finally, we demonstrate in Fig.~\ref{fig:dG-v-t} the overall structure
of spin conductance at $\epsilon_{d} = -U/2$, by varying bias voltage
and temperature.  We here scale both temperature and bias voltage by
the characteristic scale $T^{*}(\phi_{\text{so}})$.  As is seen, while
temperature always reduces spin dependence, one finds an optimal value
of bias voltage attaining the maximal amplitude of spin-dependence
around $eV \sim 0.6 T^{*}(\phi_{\text{so}})$ , as denoted by `$*$' in
Fig.~\ref{fig:dG-v-t}.

We argue that all of our numerical results are fully consistent with
the picture that this electrically generated large spin-dependent
transport can be taken as a new type of nonequilibrium Kondo effect
that is only observed at finite bias voltage: Either temperature or
bias voltage larger than the characteristic scale $T^{*}$ suppressed
the Kondo effect.  What is different from the standard (equilibrium)
Kondo effect is that we \emph{do} need finite bias voltage to support
spin moment on the dot (See \S 3).
Technically speaking, we may see the Kondo effect as the process that
strong correlation on the dot produces the density dependent
relaxation in the retarded dot Green function.  Hence if the distribution
is spin-dependent as in eq.~\eqref{eq:f-sigma}, the same Kondo mechanism
make $G^{R}_{\sigma\sigma}$ spin-dependent.
In this sense, finite bias voltage has two-sided effect: it is the
origin of the spin polarization density; yet it destroys the Kondo
effect, which is vital to produce large spin-dependent transport.
Competition of these two effects leads to the optimal bias
voltage.  

Our results are sharply contrasted with previous results of a similar
mesoscopic SOI system studied by Lu et al.~\cite{Lu07}, where
spin-dependent transport is observed more eminently around the Coulomb
blockade peaks ($n_{d} \approx 0.5, 1.5$), and larger bias voltage
induces larger spin-dependence.  Apparently the Kondo effect barely
plays an important role in their phenomena.  Indeed our estimate of
the characteristic temperature $T^{*}$ for their parameters infers
that their applied bias voltage is much larger $T^{*}$.  So we reduce
their observed spin-dependent phenomena to large nonlinear bias effect
producing spin-dependent Coulomb blockade, not to a nonequilibrium
Kondo effect and spin-dependent transport by that.

\section{Conclusion}
\label{sec:conclusion}

In conclusion, we have investigated spin dependent transport in
nonmagnetic systems through a mesoscopic spin-orbit interferometer at
low temperature.
We have shown that, even for a single-level dot, spin dependent
transport can occur electrically as a result of an intertwining effect
of the Rashba SO interaction, Coulomb interaction, and finite
bias voltage.  
We have shown that this spin-dependent transport can be well
understood as a new type of nonequilibrium Kondo effect; such
phenomenon does not appear in noninteracting dot, or in linear
conductance.  The phenomenon is suppressed either by temperature or by
bias voltage larger than the Kondo scale $T^{*}$, properly defined in
the present model.  In this regard, we view this spin-related
phenomenon as a manifestation of ``nonequilibrium strong-correlation
effect.''  Our results indicate that the interplay between the Kondo
effect and the SO coupling in the interferometer system provides a
viable option of manipulating spin degrees of freedom, especially as a
possible electrically-generated spin filtering. At the same time, we
believe that quantum dots provide a unique opportunity to study
non-equilibrium many-body effects in a well-controlled setting.

\paragraph*{Acknowledgment}

The authors appreciate T. Nemoto for helpful discussion. The work is
partially supported by Grant-in-Aid for Scientific Research (Grant
No.~22540324) from the Ministry of Education, Culture, Sports, Science
and Technology of Japan.


\bibliographystyle{jpsj}


\begin{thebibliography}{99}
\bibitem{footnote1} In the case of the
  single impurity Anderson model ($\xi\to 0$), we see $\bar
  f_{\sigma}$ equal to $(f_{L}\gamma_{L} + f_{R}\gamma_{R})/\gamma$.
\bibitem{footnote2} The definition is natural because one may write
  $\tilde{G}(0) = - (1/T^{*}) \exp(i\pi \langle n_{\sigma}\rangle)$,
  where $\langle n_{\sigma}\rangle$ is controlled by gate voltage.
\end{thebibliography}


\end{document}